\documentclass[conference]{IEEEtran}
\IEEEoverridecommandlockouts
\usepackage{cite}
\usepackage{hyperref}
\hypersetup{
    colorlinks=false,
}
\usepackage[ruled,vlined]{algorithm2e}
\usepackage{amsmath,amssymb,amsfonts,amstext}
\usepackage{algorithmic}
\usepackage{graphicx}
\usepackage{textcomp}
\usepackage{tabularx}
\usepackage{xcolor}
\def\BibTeX{{\rm B\kern-.05em{\sc i\kern-.025em b}\kern-.08em
    T\kern-.1667em\lower.7ex\hbox{E}\kern-.125emX}}
\begin{document}

\title{LSB Steganography Using Pixel\\Locator Sequence with AES\\
}

\author{\IEEEauthorblockN{$^{1}$Krishnakant Tiwari,  $^{2}$Sahil J. Gangurde}
\IEEEauthorblockA{
UG Students, Dept. Of I.T.\\
\textit{ABV - Indian Institute of Information Technology and Management, Gwalior}}
$^{1}$img\_2019030@iiitm.ac.in,  $^{2}$imt\_2019034@iiitm.ac.in\\
}

\maketitle

\begin{abstract}
 Image steganography is the art of hiding data into images. Secret data such as messages, audio, images can be hidden inside the cover image. This is mainly achieved by hiding the data into the LSB(Least Significant Bit) of the image pixels. To improve the security of steganography, this paper studied data encryption with AES(Advanced Encryption Standard) and LSB based data hiding technique with advanced user-defined encrypted data distribution in pixels other than the common linear computational method of storing data in a linear form. This data distribution file will contain the location of the data(in form of pixel numbers) to be encrypted/decrypted which is further encrypted with AES thus providing double encryption for data and its location stored over pixels. Steganography has many applications such as medical, military, copyright information, etc.\\ 
\end{abstract}


\section{Introduction}
Steganography is a  way of concealing a  file,  message, image, audio, or video into another file of the same or different category in a way such that the hidden data cannot be easily recognized by anybody other than the sender and receiver. While cryptography deals with the security of the message, steganography deals with the method of hiding the data in a way that doesn’t change the original file data at a very large cost but yet implements the secret data into it. Imperceptibility is an important feature in steganography also called transparency or anti-detection performance. The imperceptibility of steganography can be improved by enhancing the method of steganography or improving the matching relationship between secret information and carrier. Steganography complements the deficiency of concealment of the encryption. Though there is no true relationship between steganography and cryptography this paper tries to bridge the gap between the two by encrypting the secret data and also providing an alternative to a well-known steganographic method. \newline

LSB based steganography has been there for a while in which the secret data is encrypted in the LSB of a pixel in sequential order. The problem of sequential ordering of this secret data is that once the user is handed over a stego image he/she may be able to find the secret data if not encrypted with a well-known encryption algorithm. This paper tries to overcome the traditional way of implementing LSB based image steganography technique with a unique sequence that locates the pixel where the information is hidden in sequential order. The advantage of using this technique over others is that the attacker will now no longer be able to sequentially access the data as the location of the data will be scattered around in random pixel location along with AES encryption. \newline

\section{LSB substitution technique}

LSB substitution is one of the spatial domain techniques where each bit of the text or an image is substituted with the least significant bit of the original image. It is simple and easy to implement. This technique is popular because the human eye cannot easily distinguish the real image and stego image if the information change is done at least significant bit. This technique can also be extended to 2,4 or up to 8 bits but this may cause distortion and noise in the image resulting in a lossy image.\newline

To understand better, let's consider an image as a 2D matrix of pixels. Each pixel contains some value depending upon the type and depth. Consider the most widely used modes - RGB (3x8-bit pixels, true color). These values range from 0-255 (8-bit values). We can convert the message into decimal values and then into binary, by using ASCII values. Then we iterate over the pixel values one by one, after converting them to binary, we replace each least significant bit with that message bits in the sequence. To decode the message we again take the least significant bits of the pixels split them into groups of 8 and convert it back to ASCII characters to get the hidden message.\newline

\begin{figure}[htbp]
\centerline{\includegraphics{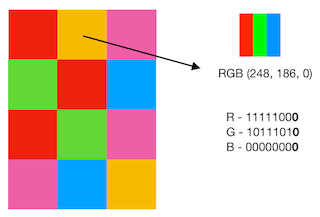}}
\caption{RGB (3x8-bit pixel) mode}
\label{LSBfig1}
\end{figure}

\begin{figure}[htbp]
\centerline{\includegraphics[width=8cm]{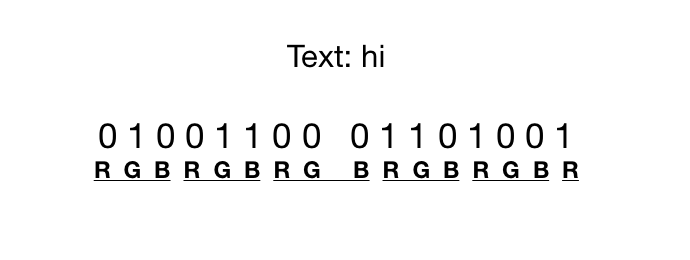}}
\caption{Encryption of string "hi" in pixels}
\label{LSBfig2}
\end{figure}

\section{AES }
For encryption purposes, we have used AES encryption.
AES is one of the most used symmetric algorithms in the world.
It's a 128-bit block cipher with three main key sizes 128,192,256 bits.
128-bit key sizes are not safe and that's why for optimum security we have used 256-bit key. AES works on bytes rather than bits. As AES has a block size of 16 bytes, these bytes are usually stored inside a 4*4 matrix.\newline
\begin{figure}[htp]
        \centerline{\includegraphics[width=4cm]{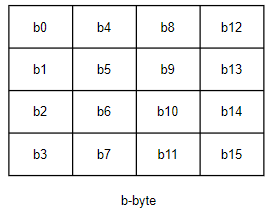}}
        \caption{Byte stored in Matrix}
        \label{fig:Byte}
    \end{figure}

Each round in AES consist of 4 layers
\begin{enumerate}
    \item SubBytes : SubBytes Layer consists of 16 S-box that are identical. Each S-box takes 1 byte of data and transform byte by taking its inverse in the finite field and then convert it using affine tranform.
    \item ShiftRows :
    Each byte is stored in a 4*4 matrix and transformed systematically.
    In the first row, no shifting is done, the 2nd row is left-shifted by one column, the 3rd row is left-shifted by 2 columns, and the 4th row by 3 columns.
    \begin{figure}[htp]
        \centerline{\includegraphics[width=8cm]{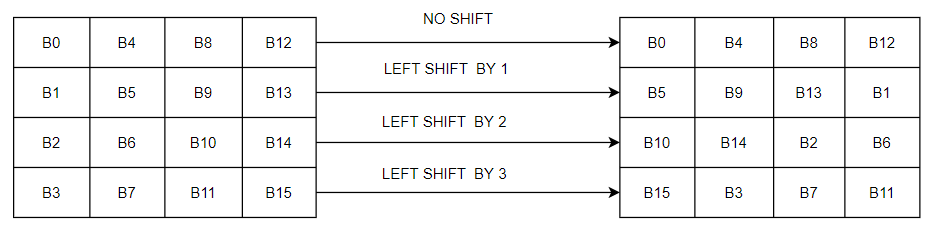}}
        \caption{Shift row operation in 4*4 matrix}
        \label{fig:Shiftrowfig1}
    \end{figure}
    \item MixColumns : In this layer first, the columns are stacked in a column matrix and then multiplied with a special 4*4 matrix. This matrix multiplication for each column results in a new 4*4 matrix which is then sent to the next round. All the operations are performed in the finite field. This helps us to distribute the changes in the whole bit pattern.
    \item AddRoundKey : The subkey which is generated for this round from key-Scheduling is added to the output from the previous layer using the XOR operation.

\end{enumerate}

\subsection{Encyption and Decryption using AES}\label{AA}
In standard implementation of AES, 14 rounds of the above operation are performed for 256-bit key size. As for each round, it requires a unique 128-bit key, AES has a key-schedule to expand a short key into various separate round keys. In the last round, the mix column layer is not present.
     \begin{figure}[htp]
        \centerline{\includegraphics[width=8cm]{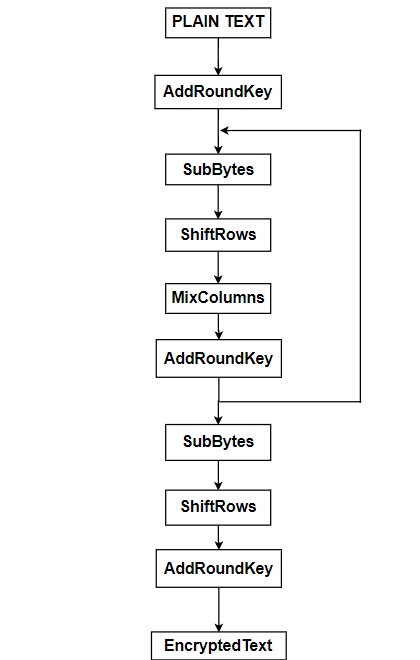}}
        \caption{Encryption in AES}
        \label{fig:RoundAes}
    \end{figure}

In the decryption process, each step is reversed by taking the inverse of all the transformations in the reverse order.\newline

Block cipher has two key property which makes them immune to an attacker i.e. confusion and diffusion. More formally confusion here means that the final output and input should have no connection between them and Diffusion means small changes in the input must have a huge effect on the final output. In AES confusion is added by the SubBytes layer and the Diffusion is added by ShiftRows and MixColumns Layer.\newline

\section{Pixel Locator Sequence} \label{PLS}
Pixel Locator Sequence (PLS) is a randomly generated pixel sequence that will generated uniquely for a particular image. PLS could also be manually created by the sender. PLS will act as a key during the decoding process.  Without PLS  decoding process is impossible. With the help of PLS, we will be able to add randomness to our encoding process.\newline

Let \textbf{N} be the total number of pixels and the length of encoded message be \textbf{N\textsubscript{enc}}. The number of pixels \textbf{N\textsubscript{p}} required to encode the given message can be simple calculated using the given formula:
\begin{equation}
    N_p = 3 \times N_{enc}
\end{equation}

Each of the 3 pixels grouped will contain the information of one character present in the encrypted text. The random distribution of pixels in PLS can be done using the \textbf{Modern Fisher-Yates Shuffle}.The following algorithm can be used to generate the required number of pixels to store the data in them.

\begin{algorithm} \label{PLSalgorithm}
\SetAlgoLined
\KwResult{Disinct \textbf{N\textsubscript{p}} numbers generated in range \textbf{0} to \textbf{N-1} stored in PLS array}
 n = N, m = N\textsubscript{p}, i = 0\;
 \While{i $\leq$ n-1}{
  arr[i] = i\;
  i++\;
 }
 i = 0\;
 \While{i $\leq$ m-1}{
  swap(arr[n-i], arr[rand()\%(n-i+1)])\;
  i++\;
 }
 PLS [], i=n-i\;
 \While{i $\geq$ n-m}{
  PLS[n-1-i] = arr[i]\;
 }
 \caption{Generating PLS sequence}
\end{algorithm}

This PLS sequence will then be used to store the data into the image at PLS[i] pixel number which is being discussed in the LSB encoding algorithm. After the encoding process, this sequence file will be transferred to an AES encrypter and will be transferred along with the stego image for the decoding process. For the manual generation of PLS user can himself put any random but distinct N\textsubscript{p} values from range 0 to N-1. 

\section{Proposed Work}
To enhance the security of LSB-based steganography what we propose is to use PLS during the encoding and decoding process of LSB so that text is hidden in a randomly distributed sequence of a pixel inside the image.
    \begin{figure}[htp]
        \centerline{\includegraphics[width=8cm]{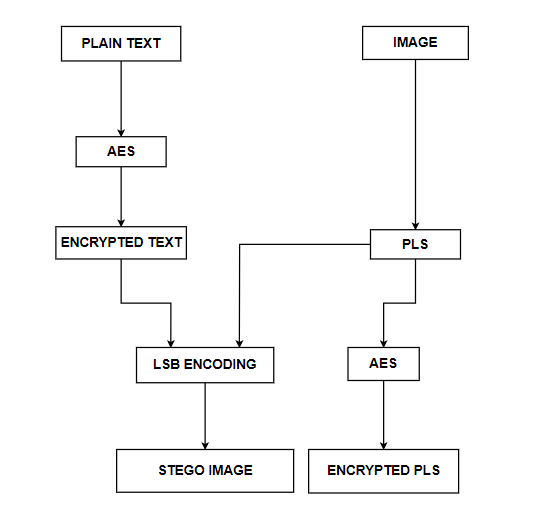}}
        \caption{Encryption Process}
        \label{PW1}
    \end{figure}

Till now, LSB based steganography is done by iterating over the pixel in systematic order. PLS will now allow us to encode the data in random order. For additional security, we will encrypt the text before encoding it into the pixel. To make the decoding possible we will have to send this sequence along with the stego image. So, we must encrypt the PLS using AES before sending it over to the receiver.
\begin{figure}[htp]
        \centerline{\includegraphics[width=8cm]{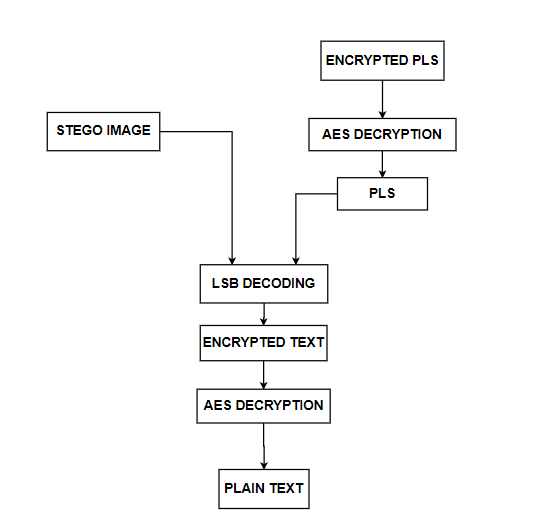}}
        \caption{Decryption Process}
        \label{PW2}
    \end{figure}
The receiver will receive a stego image along with an encrypted PLS file. At first, the PLS is decrypted using AES, and then our algorithm iterate over the PLS and from each pixel decodes the text which is hidden in the LSB of that pixel. The text which we get from this decoding is an encrypted text which further needs to be decrypted to get the original text.

\section{LSB encoding and decoding}
The PLS created will be used to distribute the data through pixels across the image. Before iterating over the image pixels we need the information of the pixel in the $i,j$ format because we consider an image to be made up of $[N] \times [M]$ matrix where N and M are whole numbers. This is easily achievable if we know $N$ and $M$. Here $N$ and $M$ represent the number of pixels in the vertical direction and horizontal directions respectively. Let $X$ be the location of the pixel from the PLS where the data has to be stored. The row and column value of $X$ can be easily calculated using \[row = X/M, column = X\%M\]

\subsection{Encoding}
Each byte of encrypted data is converted into its 8-bit binary code using ASCII values. The pixel values are read from left to right. A triad of pixels is created giving a total of nine RGB values. The RGB values of the pixel are made odd or even according to the parity of the data bits.

Let \textbf{textlist} be a list of binary string of the encrypted data  and \textbf{image} be the image matrix. Also let the number of characters in the text be \textbf{T\textsubscript{n}}. For the sake of simplicity lets assume that for every element in the \textbf{image} matrix the RGB values are easily accessible. For complete implementation refer \textit{(Algorithm 2)}. To implement this system, we used Python along with Pillow library for image manipulations.

\begin{algorithm} \label{LSBencoding}
\SetAlgoLined
\KwResult{Text encoded into the image}
 image[][], PLS[], i = 0, textlist[]\;
 \While{i $\leq$ T\textsubscript{n}-1}{
    \While{k $\leq$ textlist[i].size()}{
    text[] = textlist[i]\;
    pix [] = Append RGB values of next three pixels from PLS\;
    \While{j $\leq$ 7}{
        \eIf{text[j]=='0' and pix[j]\%2!=0}{
            pix[j]--\;
            j++\;
            }{
        {\textbf{if} text[j]=='1' and !(pix[j]\&1)\newline}
            \eIf{pix[j]==0}
            {
                pix[j]++\;
                j++\;
            }
            {
                pix[j]--\;
                j++\;
            }
        }
    }
    k++\;
    }
    i++\;
}
\caption{LSB encoding}
\end{algorithm}

\subsection{Decoding}
To decode read every 3 pixels from PLS. These 3 pixels will produce 9 different RGB values. Store these RGB values of these 3 pixels in an array. We will add '1' to the binary string if the value is odd and add '0' if the value is even. Convert this binary string into its corresponding character and append it to the final encrypted string. To decode the real message use AES to decrypt the encrypted string.

Let the \textbf{stegoimage} be the stego image matrix and \textbf{data} be an empty string where the encrypted text will be stored. For complete implementation refer \textit{(Algorithm 3)}. 

\begin{algorithm} \label{LSBdecoding}
\SetAlgoLined
\KwResult{Extracting the encrypted text from image}
 image[][], PLS[], i = 0, data = ''\;
 \While{true}{
    pix [] = Append RGB values of next three pixels from PLS\;
    binstr = ''\;
    \While{j $\leq$ 7}{
        \eIf{pix[j]\%2==0}{
            binstr += '0'\;
            }{
            binstr += '1'\;
        }
        j++\;
    }
    data += char(ASCII(binstr))\;
}
\caption{LSB decoding}
\end{algorithm}

\section{Results}
We compared five images and found the Peak Signal to Noise Ratio(PSNR) and Mean Square Error(MSE) values of them (Table 1). Generally, the steganography system should embed the secret message into the image such that the visual quality of the image is not perceptibly changed. A larger PSNR value implies lower distortion. 
\begin{equation}
PSNR = 20.\log_{10}{\frac{MAX_I}{\sqrt{MSE}}}
\end{equation}
Where MAX\textsubscript{I} is the maximum possible pixel value of that image. MSE measures the average of the square of the "error" with the error being the amount by which the estimator differs from the quantity to be estimated. Given a noise-free $m \times n$ monochrome image $I$ and its noisy approximation $K$, MSE is defined as:
\begin{equation}
MSE = \frac{1}{mn}\sum_{i=0}^{m-1} \sum_{j=0}^{n-1} [I(i,j) - K(i,j)]^2
\end{equation}
\begin{figure}[htbp]
\centerline{\includegraphics[width=8cm]{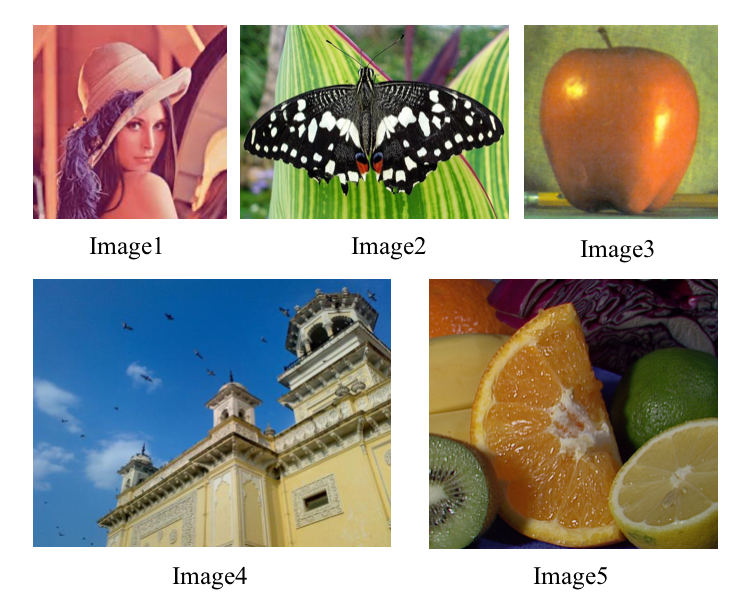}}
\caption{Sample Images}
\label{ImageComparison}
\end{figure}
Below are the sample images Fig. 8 used to derive the results. In our experimentation we tried to encode the message \textbf{'This secret message has to be embedded into the image'}. The corresponding encrypted text \textit{74890293687ebfa135e17563dc222c17696db16f6378f1a4e41e 2ae71929b41199b441f70a75b212914c0f9a64b3f3f18aec3fda5 d379238} was encoded into the image. Below in \textit{(Table 1)} we found the MSE and PSNR values of the given images. The histogram table (Fig. 9) shows the comparison between the stego and non-stego variations of the same image. The left side of the histogram table contains information about non-stego images and the right side contains information about stego images.

\begin{table}%
\begin{tabularx}{0.48\textwidth} { 
  | >{\centering\arraybackslash}X 
  | >{\centering\arraybackslash}X
  | >{\centering\arraybackslash}X
  | >{\centering\arraybackslash}X | }
 \hline
 \textbf{Images} & \textbf{Resolution} & \textbf{MSE} & \textbf{PSNR} \\
 \hline
 Image1 & 225×225 & 1.68467  & 45.86564  \\
 \hline
 Image2 & 493×356 & 2.27218  & 44.56635  \\
 \hline
 Image3 & 512×512 & 0.95186  & 48.34506  \\
 \hline
 Image4 & 512×384 & 1.19884  & 47.34317  \\
 \hline
 Image5 & 512×480 & 2.24326  & 44.62198  \\
 \hline
\end{tabularx} \newline
\caption{MSE and PSNR values of stego images}
\end{table}

\begin{figure}[htbp]
\centerline{\includegraphics[width=8cm]{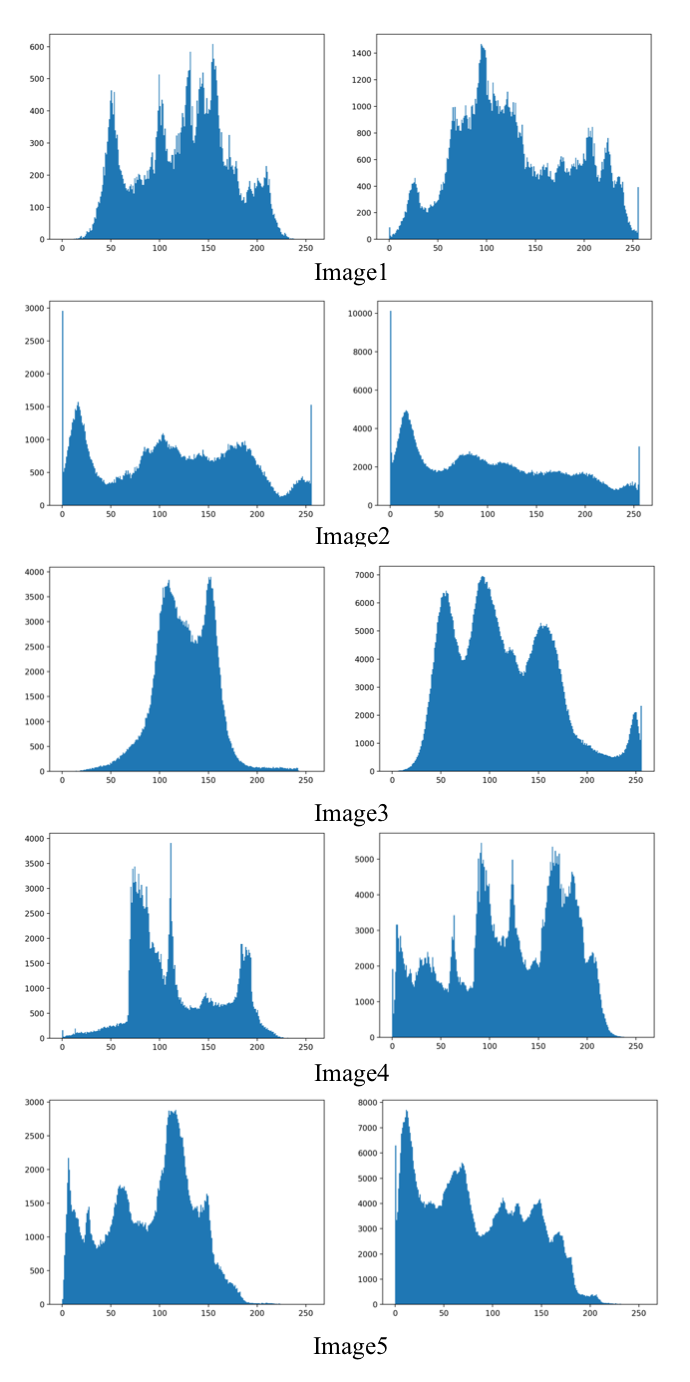}}
\caption{Stego VS non Stego images histogram}
\label{Histogram}
\end{figure}

\section{Conclusion}
In this paper, a new LSB steganography method is proposed which overcomes the security limitation of the existing LSB steganographic technique. In this method, the encoding and decoding in the image are done by using a randomly generated pixel sequence i.e. PLS which helps us to distribute data bits all over the image pixel in a pseudo-random order, making it impossible to decode without knowing the actual pixel sequence. We have also discussed PLS generation techniques using the Modern Fisher-Yates shuffling algorithm. For additional security, we have encrypted the secret text before encoding. Encoding after encryption will certainly increase the data bits of secret text in some cases but experimental results show acceptable values of PSNR and MSE.

As we have to send meta-data i.e. PLS along with the stego image, this method is not very space-efficient but we have to make that trade-off for additional security.

\end{document}